\documentclass[twocolumn,showpacs,preprintnumbers,amsmath,amssymb]{revtex4}

\usepackage{graphicx}
\usepackage{dcolumn}
\usepackage{bm}

\begin{document}

\title{Vortices in the presence of a nonmagnetic atom impurity in 2D XY ferromagnets}

\author{L. A. S. M\'ol}
 \email{lucasmol@tdnet.com.br}
\author{A. R. Pereira}%
 \email{apereira@mail.ufv.br}
\affiliation{Departamento de F\'{\i}sica, Universidade Federal de Vi\c{c}osa, 36571-000 Vi\c{c}osa, Minas Gerais, Brazil
}%

\author{A. S. T. Pires}
\affiliation{Departamento de F\'{\i}sica, Universidade Federal de Minas Gerais, Caixa Postal 702, 30123-970 Belo Horizonte, Minas Gerais, Brazil}%

\date{\today}

\begin{abstract}
Using a model of nonmagnetic impurity potential, we have examined the behavior 
of planar vortex solutions in the classical two-dimensional XY ferromagnets 
in the presence of a spin vacancy localized out of the vortex core. Our 
results show that a spinless atom impurity gives rise to an effective 
potential that repels the vortex structure.
\end{abstract}

\pacs{75.10.Hk, 75.30.Hx}

\maketitle

The easy-plane Heisenberg ferromagnet in two dimensions and continuum limit 
supports non-linear pseudo-particles with a vortex structure. These excitations 
are of paramount importance in the understanding of static and dynamical 
properties of magnetism. For example, the vortex unbinding is responsible for 
a phase transition known as Kosterlitz-Thouless transition~\cite{1}. Besides, 
these may be responsible for a central peak in the dynamical correlation 
function~\cite{2,3,4,5} observed in Monte-Carlo simulations~\cite{6,7} and experiments~\cite{8}. The simplest vortex configuration, referred as the planar vortex, occurs when 
the anisotropy is large, resulting in spin confinement to the lattice plane~\cite{9,10}.

The interaction of vortices with spatial inhomogeneities is of considerable 
importance from both the purely theoretical and applied point of view. 
Impurities and/or defects are present even in the purest of material samples 
and their effect on the motion or structure of nonlinear excitations must 
be considered when the dynamics or configurations of such solutions are 
important in the problem at hand. Recently, Zaspel, McKennan and Snaric~\cite{11} investigated, using the discrete lattice, the instability of planar vortices 
and concluded that they will be stable at a large range of anisotropies if 
there is a nonmagnetic impurity such as Cd or Zn at the center of the vortex. 
In this paper we study, using the continuum approximation, the interaction 
between a planar vortex and a nonmagnetic impurity localized out of the vortex 
center. To this end, we start defining the classical XY ferromagnet 
model, which is given by the following Hamiltonian:

\begin{eqnarray}
H=-J\sum_{m,n}(S_m^x S_n^x + S_m^y S_n^y),
\label{xy}
\end{eqnarray}

Where $J$ is a coupling constant, the classical spin vector has three components 
$\vec{S}=(S^x,S^y,S^z)$ and the summation is taking over the nearest-neighbor 
square lattice sites. This model is one of the most studied in statistical 
physics and has been found to describe a wide variety of systems with complex 
scalar order parameters, including superconductor films, Josephson junction 
arrays, and superfluid \textit{He}$^4$ films. The choice of the XY model is 
arbitrary for our purpose, since the results can be used, without modifications, 
to any other model with XY symmetry such as the classical easy-plane 
ferromagnetic model. It is convenient to parametrize the spin field in terms of spherical 
coordinates as follows: $\vec{S}=S(\cos\,\theta \cos\,\phi, \cos\,\theta \sin\,\phi, \sin\,\theta).$ By a straightforward generalization of arguments used to obtain the continuum limit of Heisenberg Hamiltonians in the case of one dimension~\cite{12},
we can write the continuum version of the Hamiltonian (\ref{xy}) as

\begin{equation}
H=\frac{J}{2}\int d^2r\left[\frac{m^2(\vec{\nabla }m)^2}{1-m^2}+(1-m^2)(\vec{\nabla }\Phi)^2+\frac{4}{a_0^2}m^2\right],
\end{equation}

where $m=\sin\,\theta$, $a_0$ is the lattice constant and we have taken 
$S^2=1$. One can obtain the motion equations, $\partial m/\partial t=\delta H / \delta \Phi, \quad \partial \Phi /\partial t = -\delta H / \delta m,$ 
for this field theory in the usual way using the pair of canonically 
conjugated variables $m$ and $\Phi$.

Now, consider that the system contains a nonmagnetic impurity concentration in the plane. For simplicity, we consider only one nonmagnetic atom present. If we remove a spin from the lattice, the nearest neighbors of that spin will have a coordination number of three, instead of bulk-spin coordination number of four. Therefore, such boundary spins would have larger fluctuations than the bulk spins and it is conceivable that the non-linear configurations such as a vortex, would preferentially nucleate around this vacancy. In fact, in this circumstance, the vortex energy is lowered, since the nonmagnetic impurity at the vortex center will remove the nearest-neighbor exchange bonds at the impurity in a radially symmetric way without modifying the symmetric configuration of the vortex, while its energy in the region without the impurity remains the same. However, the vortex energy increases logarithmically with the system size $L$, and in an infinitely extended system, this energy would diverges as $L\rightarrow \infty $ so that we should not expect that a single vortex could nucleate around the spin vacancy. In fact, vortices are always created in pairs of vortex-antivortex having finite energy and separated by a few lattice constants. Then, if one member of the pair nucleates around the impurity, the other member will be near it. Since a vortex pair does not have a cylindrical symmetry, the energy of this system does not necessarily decrease, although there are less nearest-neighbor exchange bounds, because 
the spin vacancy may deform the pair configuration, increasing its energy. 
Nevertheless, the interest of this paper is to study the behavior of a vortex in the system, that is not nucleated at the position of the vacancy. In this case things must change considerably because the spin vacancy may also deform the single vortex configuration. To take into account the nonmagnetic impurities we consider the following modified XY ferromagnetic Hamiltonian in the continuum limit

\begin{eqnarray}
H_I=\frac{J}{2}\int d^2r\left[\frac{m^2(\vec{\nabla }m)^2}{1-m^2}+(1-m^2)(\vec{\nabla }\Phi)^2+\frac{4}{a_0^2}m^2\right]\nonumber \\ \times V(\vec{r}),
\label{h_imp}
\end{eqnarray}

where $V(\vec{r})$ is a nonmagnetic impurity potential given by

\begin{eqnarray}
V(\vec{r})=\left\{\begin{array}{rc}
1&\mbox{if}\quad |\vec{r}-\vec{r}_0|\geq b,\\
0&\mbox{if}\quad |\vec{r}-\vec{r}_0|< b.
\end{array}\right.
\label{def_v}
\end{eqnarray}

Here the impurity is centralized at the point $\vec{r}_0$ and has the form of a circle with diameter equal $2b$. There is a circular region in the plane, around the point $\vec{r}_0$, without any magnetic interaction.

Substituting Eq. (\ref{def_v}) into the equations of motion we get

\begin{eqnarray}
\frac{1}{J}\frac{\partial \theta }{\partial t}=\cos\, \theta V(\vec{r}) \nabla ^2 \Phi -2\sin\,\theta V(\vec{r})\vec{\nabla }\theta\cdot \vec{\nabla}\Phi \nonumber\\ + \cos\,\theta \vec{\nabla}V(\vec{r}) \cdot \vec{\nabla}\Phi,
\label{eq_mov_1}
\end{eqnarray}

\begin{eqnarray}
\frac{1}{J}\frac{\partial \Phi }{\partial t}=-\tan \theta \sin \theta V(\vec{r}) \nabla ^2 \theta - \sin \theta V(\vec{r}) (\vec{\nabla} \theta)^2 \nonumber \\+ \sin \theta V(\vec{r})[4/a_0^2 - (\vec{\nabla}\Phi)^2]
-\tan ^2 \theta \sin \theta \vec{\nabla}V(\vec{r})\cdot \vec{\nabla} \theta.
\label{eq_mov_2}
\end{eqnarray}

Our interest is in planar and static solutions in the presence of this 
nonmagnetic impurity potential. Hence, we take $\partial \theta /\partial t=\partial \Phi /\partial t=0$ and $m=\sin \theta=0$ in Eqs. (\ref{eq_mov_1}) and (\ref{eq_mov_2}), obtaining only one and simpler equation to be solved

\begin{eqnarray}
V(\vec{r})\nabla ^2 \Phi=-\vec{\nabla}V(\vec{r})\cdot\vec{\nabla}\Phi.
\label{eq_mov_est}
\end{eqnarray}

One point to note in this equation is its dependence on the spin field around the position of the impurity. If the vacancy is localized in a region where the spin configuration consists of aligned spins like a domain with all spins aligned along some direction $(|\vec{\nabla}\Phi |\approx 0)$, the spin field practically does not feel the presence of the impurity. However, an impurity placed in a region where the spin directions vary considerably $(|\vec{\nabla}\Phi|>>1/a_0)$ may have a strong coupling with the spin field and may modify the initial spin configuration for large distances.

In polar coordinates the vectors $\vec{r}$ and $\vec{r}_0$ are written as 
$(r, \phi),$ and $(r_0,\phi _0)$ respectively. To solve Eq.(7), we note first that the gradient of the impurity potential can be expressed as 

\begin{eqnarray}
\vec{\nabla} V(\vec{r})=a_0 [\hat{r}\cos (\alpha -|\phi -\phi _0|)+\hat{\phi}\sin(\alpha -|\phi-\phi _0|)]\nonumber\\ \times\delta (\vec{r}-\vec{r}_0-\vec{b}),
\label{grad_v_1}
\end{eqnarray}

where $\delta$ is the Dirac delta function and $\alpha$ is the angle that the vector $\vec{b}$, with origin at the point $\vec{r}_0$ and end at a point on the circumference of the potential, makes with the vector $\vec{r}_0$. As we are interested in a local impurity with atomic dimensions, we make $b\rightarrow 0$ in the continuum limit (to be more precise, we should make $b\rightarrow a_0$) indicating that the impurity is an atom (such as Zn, Mg or Cd for example). In this case, $\vec{r}\rightarrow \vec{r}_0, \quad \phi \rightarrow \phi_0,$ 
and we rewrite Eq. (\ref{grad_v_1}) as

\begin{eqnarray}
\vec{\nabla}V(\vec{r})\approx a_0[\hat{r} \cos(\alpha)+\hat{\phi} \sin(\alpha)]\delta(\vec{r}-\vec{r}_0),
\label{grad_v_2}
\end{eqnarray}

where we can interpret $\cos(\alpha)$ and $\sin(\alpha)$ as anisotropic 
coupling constants. This coupling depends on the direction one looks, if the observer center is placed on the impurity position.

Considering Eq. (\ref{eq_mov_est}) with $V(\vec{r})=1$ at the left side (this fails only at the point $\vec{r}_0$, since the impurity is local) and supposing that the vortex structure is modified by the presence of the nonmagnetic impurity, we write $\Phi=\Phi _0 + \Phi _1$, where $\Phi _0 = \arctan (y/x)$ is the traditional single vortex solution for a vortex with center localized at the origin and $\Phi _1$ is the deformation caused by the spinless impurity localized at $\vec{r}_0$. Thus, Eq. (\ref{eq_mov_est}) with the above considerations can be written as

\begin{eqnarray}
\nabla ^2 (\Phi _0 + \Phi _1)=-a_0\vec{\nabla}(\Phi_0+\Phi_1) \nonumber\\ \cdot[\hat{r}\cos(\alpha)+\hat{\phi}\sin(\alpha)]\delta(\vec{r}-\vec{r}_0)
\label{lap_p0+p1}
\end{eqnarray}

Using the fact that $\nabla^2 \Phi_0 =0 $ and taking $\vec{\nabla}(\Phi_0+\Phi_1)\cong \vec{\nabla}\Phi_0=\frac{1}{r}\hat{\phi}$ 
near the point $\vec{r}_0$, Eq. (\ref{lap_p0+p1}) can then be approximated by

\begin{eqnarray}
\nabla^2 \Phi_1 = -\frac{a_0}{r_0} \sin (\alpha) \delta (\vec{r}-\vec{r}_0),
\end{eqnarray}

or

\begin{eqnarray}
\nabla^2 \left[\frac{-2 \pi r_0 \Phi_1}{a_0 \sin(\alpha)}\right]=2\pi \delta(\vec{r}-\vec{r}_0).
\end{eqnarray}

This is easily solved using the fact that in two dimensions, $\nabla^2 \ln(r)=2\pi \delta(\vec{r})$. We get 

\begin{eqnarray}
\Phi_1(\vec{r})=\frac{a_0 \sin(\alpha)}{2 \pi r_0}\ln \left(\frac{|\vec{r}-\vec{r}_0|}{a_0}\right).
\end{eqnarray}

Writing the anisotropic coupling constant along the $\alpha$-direction in 
terms of $r$, $\phi$, the vortex structure with its center at the origin in the presence of a nonmagnetic impurity localized at $\vec{r}_0$ is given by

\begin{eqnarray}
\Phi=\arctan(y/x) - \frac{a_0}{2 \pi r_0}\frac{r \sin (\phi - \phi_0)}{|\vec{r}-\vec{r}_0|}\ln \left(\frac{|\vec{r}-\vec{r}_0|}{a_0}\right).
\label{sol}
\end{eqnarray}

The configuration of this deformed vortex is shown in figures 1 and 2. Although the continuum theory cannot be applied near the vortex core, in Fig. 1 we have considered the impurity one lattice sapcing from the vortex center just to emphasize the vortex deformation as the vortex core approaches the impurity.We notice that if $r_0$ is large (the vortex center is far away from the spin vacancy) the vortex practically keeps the same original form, but for small $r_0$ the vortex configuration suffers a severe modification, mainly in the region in which the impurity is located. This is due to the fact that the gradient of the spin field is small in the region of the impurity if it is far way from the vortex center and large in the region of the impurity if it is near the vortex center. When animpurity is near the vortex core, Eq. (\ref{sol}) implies (as it can also be seen partially in figure 1) that a large domain with all spins aligned along the direction perpendicular to $\vec{r}_0$ will be formed in a region located after the impurity position $\vec{r}_0$.

In order to calculate the energy of this planar solution we consider the 
Hamiltonian (\ref{h_imp}) with $m=0$, obtaining $E_I=\int(\vec{\nabla}\Phi)^2V(\vec{r})d^2r$. As we have seen, the field $\Phi$ describes a single vortex at the origin in the presence of an impurity at distance $r_0$ away. The effective potential experienced between the two defects (one defect in the spin field and the other 
in the lattice structure) is defined as

\begin{eqnarray}
U_{eff}(r_0)=E_I-E_\nu,
\end{eqnarray}

where $E_\nu = \pi J \ln(L/a_0)$ is the energy of a single vortex in the 
absence of impurities. Making suitable approximations, we find that such 
effective potential results in a repulsive central interaction with a 
dominant term given by

\begin{eqnarray}
U_{eff}(r_0)\cong \frac{a_0^2 E_\nu^3}{24 \pi^4 J^2}\frac{1}{r_0^2}.
\label{res}
\end{eqnarray}

We see, therefore, that the presence of a nonmagnetic impurity increases the vortex energy as the distance between the impurity and the vortex decreases. In a ferromagnet with a size of the order of $L\approx 10^8a_0$ ( a few centimeters), a spinless atom impurity situated about $2a_0$ from the vortex core would increase the vortex energy about 36\%. Note that the effective potential barrier becomes infinity as $r_0\rightarrow 0$ and it is energetically favorable that vortices and impurities become far apart. But, if the calculations were taken considering that the spin vacancy is localized at the vortex center, we would have $\vec{\nabla}V(\vec{r})=a_0 \delta(\vec{r})\hat{r}$ and near the vortex core $\vec{\nabla}\Phi \approx (1/a_0)\hat{\phi}$, leading to $V(\vec{r})\nabla^2\Phi=0$. As a consequence, in the region without the spinless impurity, where $\nabla^2\Phi=0$, one gets the same typical solutions and the vortex structure does not suffer any alteration. Hence, the only one effect 
of a central nonmagnetic impurity, is to make the vortex energy to decrease, because of the nonexitence of nearest-neighbor exchange bonds at the impurity. Nevertheless, as we suggested earlier, a single vortex with infinite energy may not nucleate by itself around the impurity, since these are created in pairs. Besides, Eq. (\ref{res}) shows that an infinite potential barrier has to be exceeded by the vortex core in order that it might reach the nonmagnetic impurity and the minimum of energy. Then, one should not expect to find a single vortex with a spin vacancy localized in its center.

In Summary, vortices prefer to stay far way from nonmagnetic impurities and 
hence, the spin dynamics must be affected by these lattice defects. Our 
calculations could also be taken for two-dimensional easy-plane antiferromagnets. It would be carried out in essentially the same way, leading to similar results. The structure and motion of vortices in two-dimensional magnets may be driven by the presence of spinless impurities due to the repulsive effective potential. Since the dynamical structure factor is the Fourier transform of the vortex spatial and temporal configuration, we expect that nonmagnetic impurities may cause changes in the central peak~\cite{2,3,4,5} and also in the electron paramagnetic resonance linewidth~\cite{13}, which must be seen in neutron scattering and resonance experiments. However, much work has to be done in order to see these effects. Besides, since the vortex energy is modified, the Kosterlitz-Thouless temperature may also be affected by the presence of impurities. In fact this theory holds also for temperatures below the Kosterlitz-Thouless temperature $T_{KT}$, where vortices are bound in pairs. However, the problem of vortex pairs interacting with nonmagnetic impurities and its influences on $T_{KT}$ will be treated in a future paper.We also suggest that the above calculations may have some relevance to high-T$_c$ superconductors, because a common feature of all high-transition-temperature cuprates is the proximity between antiferromagnetic 
and d-wave superconducting phases controlled by the doping. The effect of impurities on superconductors has been of theoretical and experimental interest even in its own right for a long time. Recent nuclear magnetic resonance measurements have shown that when a $Cu^{2+}$ in the $Cu-O$ plane is substituted by a strong nonmagnetic impurity, such as $Zn^{2+}$, an effective magnetic moment can be induced on the $Cu$ sites around the impurity site~\cite{15,16}. The physical picture implied by these experiments is that antiferromagnetic correlations are enhanced, not destroyed, around impurities in these cuprates.

\begin{acknowledgments}
This work was partially supported by CNPq and FAPEMIG (Brazil).
\end{acknowledgments}

\begin{figure*}
\includegraphics[height=13cm, keepaspectratio]{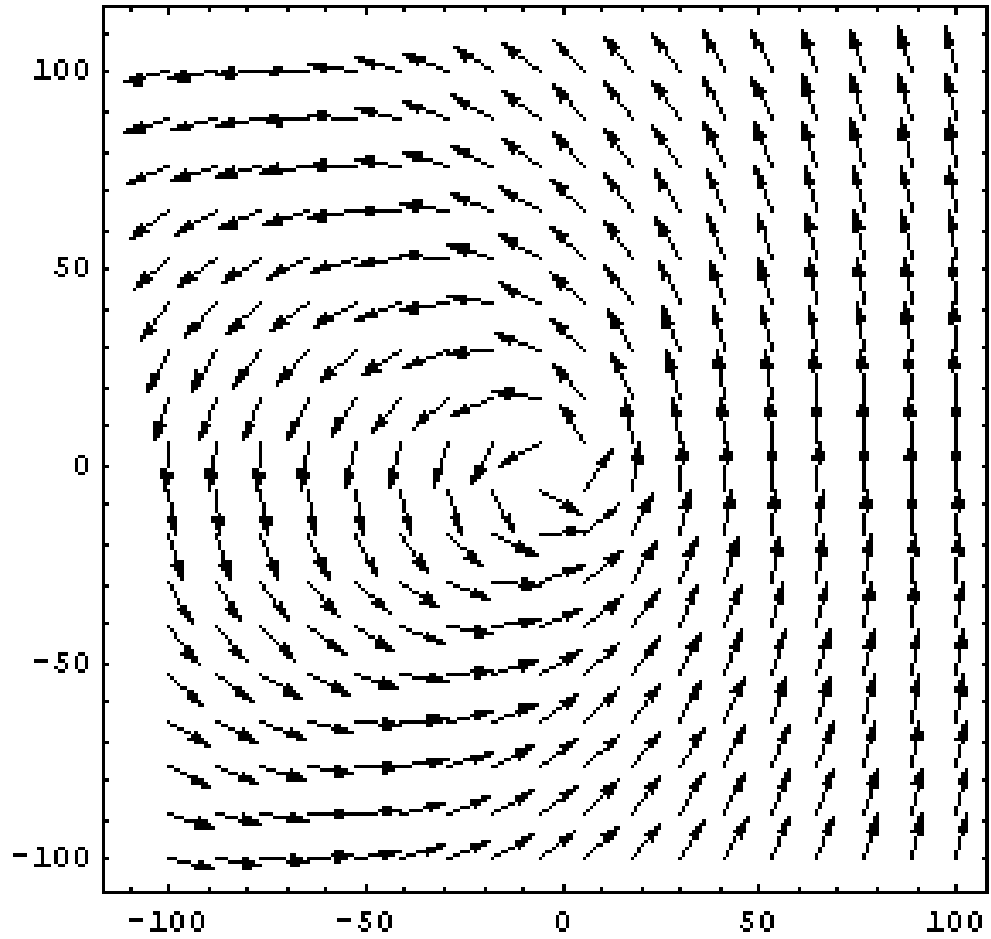}
\caption{Structure of a single vortex with center at (0,0) in the presence of a 
nonmagnetic impurity located at the site (1,0). Since the nonmagnetic impurity 
is near the vortex center, the vortex experiences a strong effect of the spinless 
atom impurity.}
\end{figure*}

\begin{figure*}
\includegraphics[height=13cm, keepaspectratio]{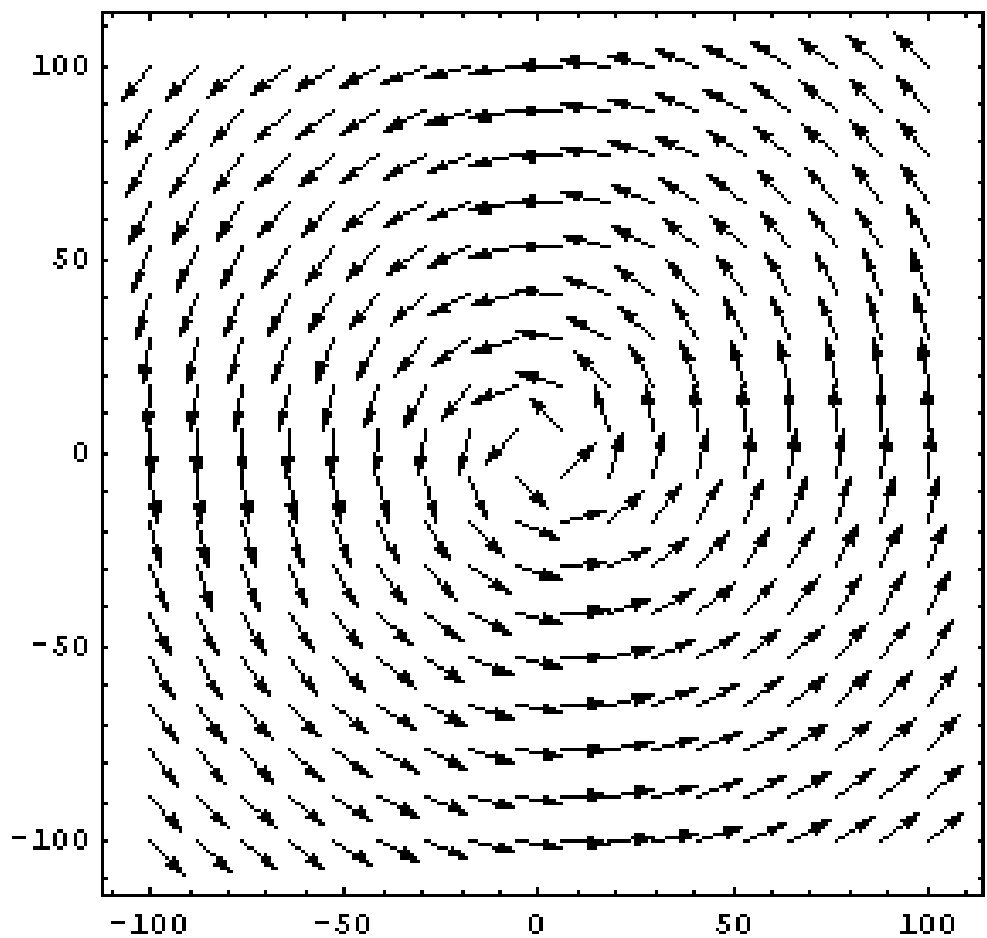}
\caption{Here, the impurity is located at (5,0). Note that the vortex structure 
is almost perfect, indicating that a vacancy put away from the vortex core has 
small influence on the vortex structure.}
\end{figure*}

\end{document}